# GPCALMA: A Tool For Mammography With A GRID-Connected Distributed Database

U. Bottigli[1], P. Cerello[2], S. Cheran[3], P. Delogu[4], M.E. Fantacci[4], F. Fauci[5], B. Golosio[1], A. Lauria[6], E. Lopez Torres [7], R. Magro[5], G.L. Masala[1], P. Oliva[1], R. Palmiero[6], G. Raso[5], A. Retico[4], S. Stumbo[1], S. Tangaro[8]

[1]*Università di Sassari and Sezione INFN di Cagliari, Italy*
[2]*Sezione INFN di Torino, Italy*
[3]*University of Siegen, Germany*
[4]*Università and Sezione INFN di Pisa, Italy*
[5]*Università di Palermo and Sezione INFN di Catania, Italy*
[6]*Università "Federico II" and Sezione INFN di Napoli, Italy*
[7]*CEADEN, Havana, Cuba*
[8]*Università di Bari and Sezione INFN di Cagliari, Italy*

**Abstract.** The GPCALMA (Grid Platform for Computer Assisted Library for MAmmography) collaboration involves several departments of physics, INFN (National Institute of Nuclear Physics) sections, and italian hospitals. The aim of this collaboration is developing a tool that can help radiologists in early detection of breast cancer. GPCALMA has built a large distributed database of digitised mammographic images (about 5500 images corresponding to 1650 patients) and developed a CAD (Computer Aided Detection) software which is integrated in a station that can also be used for acquire new images, as archive and to perform statistical analysis. The images (18x24 $cm^2$, digitised by a CCD linear scanner with a 85 μm pitch and 4096 gray levels) are completely described: pathological ones have a consistent characterization with radiologist's diagnosis and histological data, non pathological ones correspond to patients with a follow up at least three years.
The distributed database is realized throught the connection of all the hospitals and research centers in GRID tecnology. In each hospital local patients digital images are stored in the local database. Using GRID connection, GPCALMA will allow each node to work on distributed database data as well as local database data.
Using its database the GPCALMA tools perform several analysis. A texture analysis, i.e. an automated classification on adipose, dense or glandular texture, can be provided by the system. GPCALMA software also allows classification of pathological features, in particular massive lesions (both opacities and spiculated lesions) analysis and microcalcification clusters analysis. The detection of pathological features is made using neural network software that provides a selection of areas showing a given "suspicion level" of lesion occurrence.
The performance of the GPCALMA system will be presented in terms of the ROC (Receiver Operating Characteristic) curves. The results of GPCALMA system as "second reader" will also be presented.

# INTRODUCTION

Early diagnosis of breast cancer in asymptomatic women makes possible the reduction of breast cancer mortality [1]. At this moment an early diagnosis is possible thanks to screening programs, which consist in a mammographic examination performed for 49-69 years old women. It has been estimated that screening programs radiologists fail to detect up to approximately 25% breast cancers visible on retrospective reviews and that this percentage increases if minimal signs are considered [2,3]. Sensitivity (percentage of pathologic images correctly classified) and specificity (percentage of non pathologic images correctly classified) of this examination increase if the images are analysed independently by two radiologists [4]. So independent double reading is now strongly recommended as it allows to reduce the rate of false negative examinations by 5-15% [5,6]. Recent technological progress has allowed to develop a number of Computer Aided Detection (CADe) systems [7]. We report here a description of the algorithms currently used by GPCALMA CADe and their performance in the automated search of pathological masses and microcalcification clusters. We report also a comparison between our CADe and a commercial one.

# DESCRIPTION OF A GPCALMA STATION

The hardware requirements for the GPCALMA CAD Station are very simple: a PC with SCSI bus connected to a planar scanner. Given the size of information to store big disks and a CD-ROM recorder are recommended. A high resolution monitor helps the human radiological diagnosis. The station can process mammograms directly acquired by the scanner and/or images from file and allows human and/or automatic analysis of the digital mammogram. The images (18x24 $cm^2$) have been digitised by a CCD linear scanner with a 85 μm pitch and 4096 gray levels and are stored in 10.5 Mbytes data files. The human analysis produces a diagnosis of the breast lesions in terms of kind, localization on the image, average dimensions and, if present, histological type. The automatic procedure find the regions of interest (ROIs) on the image which have a chance (greater than the chosen threshold value) of containing an interesting area. For each image it is possible to insert or modify diagnosis and annotations, manually select the ROIs corresponding to the radiologists geometrical indication. There is also an interactive windows which allows zoom, windowing, gray levels and contrast selection. The operator can also start the CADe analysis by choosing in other interactive windows the threshold for ROIs selection for microcalcifications or opacities. The station allows also for queries and statistical studies on the local database.

Figure 1 shows an example in which the green circle indicates the the ROI indicated by the radiologist as suspected for the presence of spiculated lesions with granular microcalcifications. The CADe has correctly indicated two ROIs suspect for the presence of pathological masses (red circles, threshold=0.9) and two ROIs suspect for the presence of microcalcifications (red rectangles, threshold=0.95). The histological examination has confirmed this detection, indicating the presence of a ductal infiltrating carcinoma with granular microcalcifications.

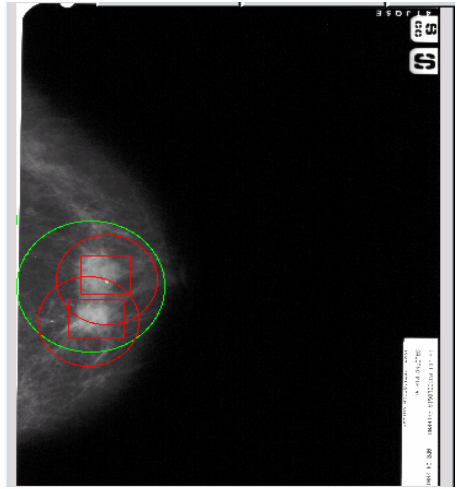

**Figure 1:** An example from the GPCALMA CADe results.

## GPCALMA CADe SOFTWARE

### Opacities and Spiculated Lesions

Masses (fig. 2) are rather large objects with very different shapes and show up with a faint contrast slowly increasing with time. In GPCALMA database, the mean diameter of such lesions, as indicated by our radiologists, is 2.1cm. We have developed algorithms for recognition of opacities in general and specifically for spiculated lesions, which present a particular spike shape.

The interesting areas are selected by the construction of a structure with concentric rings centered on local intensity maxima until the mean pixel value reaches a fixed threshold, thus identifying ROIs consisting of circles of radius R. As a further step, for searching spiculated lesions, a spiral is unrolled around each maximum. For opacities, features are extracted by calculating the average intensity, variance and skewness (index of asymmetric distribution) of the pixel value distributions in circles of radius 1/3 R, 2/3 R and R, respectively. In the case of spiculated lesions, the number of oscillations per turn is calculated and processed by means of a Fourier Transform.

The features so extracted are used as input of a feed-forward neural network which perform the final classification. This network has an output neuron whose threshold (i.e. a number comprised between 0 and 1) represents the of suspiciousness of the corresponding ROI.

### Microcalcifications clusters

A microcalcification is a rather small (0.1 to 1.0 mm in diameter) but very brilliant object. Some of them, either grouped in cluster (fig. 2) or isolated, may indicate the presence of a cancer. In the GPCALMA database, the mean diameter of microcalcification clusters, as indicated by our radiologists, is 2.3cm.

Microcalcification cluster analysis was made using the following approach:
- the digital mammogram is divided into 60x60 pixels wide windows;
- the windows outside the breast image are rejected;
- the windows are statistically selected comparing the local and the global maxima;
- the windows are shrunk from 60x60 to 7x7 and are classified (with or without microcalcifications clusters) using a FFNN with 49 input, 6 hidden, and 2 output neurons;
- the windows are processed by a convolution filter to reduce the large structures;
- a self-organizing map (a Sanger's neural network) analyzes each window and produces 8 principal components;
- the principal components are used as input of a FFNN able to classify the windows matched to a threshold (the response of the output neuron of the neural network);
- the windows are sorted by the threshold;
- at maximum three windows are memorized, if its threshold exceeds a given value;
- the selected windows are zoomed to 180x180 pixels, i.e. 15x15 mm$^2$;
- the overlapping windows are clusterized.

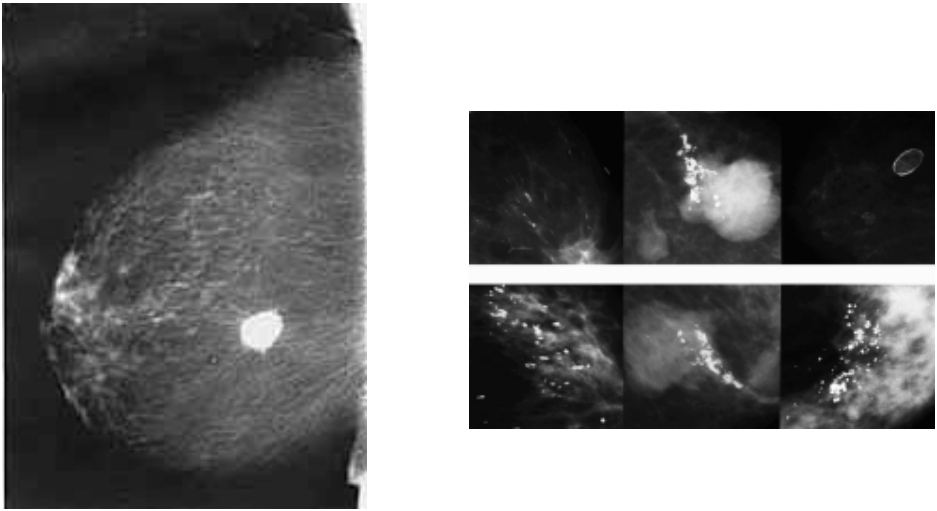

**Figure 2:** Some examples of massive lesion (left) and microcalcification clusters (right).

# RESULTS

## Opacities and Spiculated Lesions

The FFNN previously described has been trained with a training set of 515 images (102 containing opacities, 413 without) and tested on a test set, composed of other 515 images (again 102 containing opacities and 413 without. The results of such a classification are reported in the ROC (Receiver Operating Characteristics) curve of figure 3, in which are plotted sensitivity (true positives fraction) versus 1-specificity (false positives fraction) for different threshold values. The best results are 94% for sensitivity and 95% for specificity.

## Microcalcifications clusters

The previously described procedure has been used on a dataset of 676 images containing microcalcifitations and 995 images without microcalcification clusters. In particular, 370 images with microcalcifications clusters and 495 without have been utilized in the training phase and 306 with and 500 without for the test. The results obtained in this test are summarized in the ROC (Receiver Operating Characteristic) curve reported in figure 3, realized varying the threshold by which the interesting windows are sorted. The best results are 92% for both sensitivity and specificity.

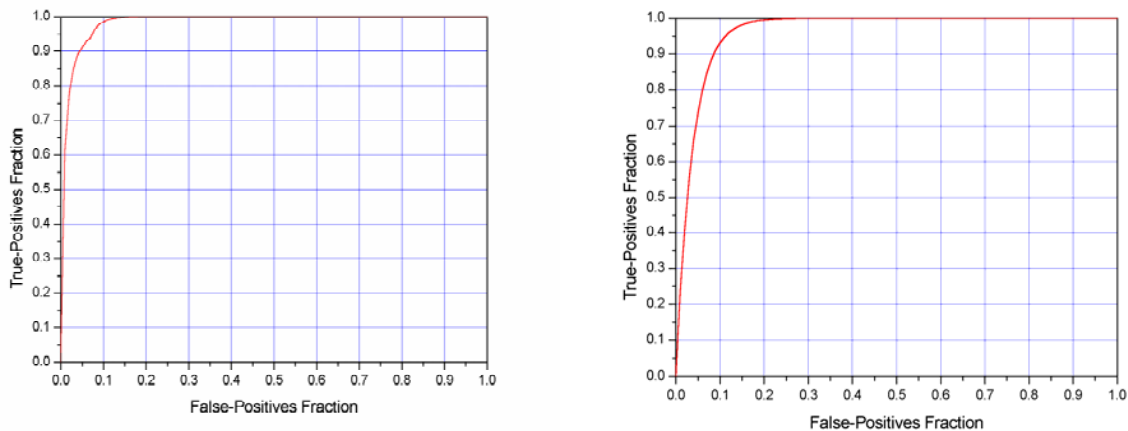

**Figure 3:** ROC curves for opacities and spikes (left, made on of 515 images, 102 containing opacities) and microcalcification (right made on of 865 images, 370 containing microcalcification clusters).

## COMPARISON WITH A COMMERCIAL CAD SYSTEM

CADX Second Look is a commercial CADe system. It has been tested for micorcalcification analysis on 98 images coming from the GPCALMA database. The results are a sensitivity of 90% and a specificity of 70%. The same sensitivity is obtained by the GPCALMA CADe in conjunction with a specificity of 91%, while the same specificity is obtained in conjunction with a sensitivity of 99.9%.

In table 1 and 2 the results of the comparison of GPCALMA and Second Look, used as second readers, are shown. Data are obtained with a data set of 190 (70 pathological, 120 without lesions) images and they are produced by 3 different radiologists (A,B,C), with a different degree of experience.

| TABLE 1: Sensitivity (and confidence interval) | | | |
|---|---|---|---|
| | Alone (C.I.) | With CADx (C.I.) | With GPCALMA (C.I.) |
| A | 82.8% (4.5%) | 94.3% (2.8%) | 94.3% (2.8%) |
| B | 80.0% (4.8%) | 88.2% (3.8%) | 90.0% (3.6%) |
| C | 71.5% (5.4%) | 82.9% (4.5%) | 87.1% (4.0%) |

| TABLE 2: Specificity (and confidence interval) | | | |
|---|---|---|---|
| | Alone (C.I.) | With CADx (C.I.) | With GPCALMA (C.I.) |
| A | 87.5% (3.0%) | 84.2% (3.3%) | 87.5% (3.0%) |
| B | 91.7% (2.6%) | 85.9% (3.2%) | 88.4% (2.9%) |
| C | 74.2% (4.0%) | 70.8% (4.2%) | 70.9% (4.1%) |

## GRID CONNECTION OF DISTRIBUTED DATABASE

GPCALMA database is made of about 5500 images completely described: pathological ones have a consistent characterisation with radiologist's diagnosis and histological data, non pathological ones correspond to patients with a follow up at least three years. The database is distributed over various Italian hospitals, that will be connected using GRID technologies, allowing each node to work on the whole database. Each image fits to 10.5 Mbytes and the transfer of all patient images from a local database to another place could require a large amount of time, so the images transfer is not suitable. The data of each patient are shown to the user as if they were stored into a large central database alone rather than into a shared database. This is possible by using the metadata abstracting information. The GRID philosophy in mammographic CADe is that it's better move code rather data. Therefore the user has the advantage to work using the CADe on the remote patient images without the transfer of the images and without knowing their real location.